\documentclass[french]{article-hermes}
\usepackage{epsfig}
\begin{document}

\newcommand{\SNF}{service technique}
\newcommand{\SNFs}{services techniques}
\newcommand{\QoS}{qualité de service}
\newcommand{\ps}{P1S}
\newcommand{\pss}{P1Ss}
\newcommand{\pe}{L2S}
\newcommand{\pes}{L2Ss}

\title[Gestion dynamique des services techniques]{Gestion Dynamique des Services Techniques pour Modèle à Composants}
\author{Colombe Hérault \andauthor Sylvain Lecomte}
\address{
    LAMIH / ROI / SID, UMR CNRS 8530\\
    Université de Valenciennes - Le Mont Houy,\\
    59313 Valenciennes Cedex 9 - France\\
    [3pt] \{colombe.herault, sylvain.lecomte\}@univ-valenciennes.fr}

\resume{Les nouvelles applications étant destinées à des environnements de plus en plus hétérogènes, il est
indispensable de proposer des solutions de développement qui répondent au mieux aux besoins d'adaptation des
nouveaux services. La programmation par composants répond en partie à cela, en permettant d'interchanger les
briques logiciels afin de fournir la version d'un composant la plus adaptée à son contexte d'exécution.
Néanmoins, la plupart des implantation industrielles des modèles à composants ne permettent pas de fournir aux
composants les services techniques (nommage, courtage, sécurité, transaction, etc.) les mieux adaptés. Dans cet
article, nous proposons de définir les services techniques eux-mêmes sous forme de composants. Nous détaillerons
notre proposition, en la basant sur le modèle à composants Fractal de Objectweb. Puis, nous apportons des
solutions pour l'utilisation de ces nouveaux services techniques à base de composants et proposons un ensemble
de composants de gestion qui permettent de gérer de façon dynamique et automatique les composants obtenus. Enfin
nous présentons le prototype du système proposé.} \abstract{The new applications being intended for more and
more heterogeneous environments, it is necessary to propose solutions of development which answer in best the
necessities of adaptation of new services. Component-based programming partially answers this aim, allowing easy
replacement of software blocks in order to provide the most adapted version of a component. Nevertheless, most
of the industrial component-based model implementations do not allow to provide to components the most adapted
technical services (naming, trading, security, transaction, etc.). In this paper, we suggest defining technical
services themselves under the shape of components. We shall detail our proposition, by basing it on the Fractal
component model of Objectweb. Then, we shall bring solutions for the use of these new component-based technical
services and shall propose a set of management components which allow to administer in a dynamic and stand-alone
way the obtained components. Finally we present the prototype of the proposed solution.} \motscles{services
techniques, adaptabilité, architecture à composants, intergiciel.} \keywords{technical services, adaptability,
component-based architecture, middleware.} 

\proceedings{DECOR'04, Déploiement et (Re)Configuration de Logiciels}{135}

\maketitlepage
\section{\label{section1}Introduction}
L'émergence des intergiciels fut en grande partie motivée par le besoin d'interopérabilité dans un contexte
fortement distribué entre des modules logiciels écrits à des époques différentes avec des langages hétérogènes.
Ensuite on a voulu apporter une certaine qualité de service à ces modules logiciels et une plus grande facilité
de développement et de réutilisabilité. L'une des solutions adoptées fut de séparer code fonctionnel et
non-fonctionnel. Ainsi le développeur se focalise sur la logique applicative et délègue l'implantation des
services techniques (ou non-fonctionnels) à la plate-forme d'exécution. Cependant, les terminaux participant aux
applications distribuées sont de plus en plus hétérogènes: de l'ordinateur très puissant aux cartes à
microprocesseur. A cause de la diversité de l'offre de terminaux, le programmeur doit assurer la portabilité de
ses applications. Une solution pour gérer cette hétérogénéité au niveau de l'application est d'utiliser le
modèle de programmation par composants \cite{art30}. Cependant, les modèles à composants actuels ne permettent
pas d'adapter les \SNFs\ à l'environnement d'exécution: le développeur peut uniquement décider d'utiliser ou non
un service. On ne lui offre pas le choix du modèle implantant le code non-fonctionnel. Les composants qu'il
écrit utilisent donc toujours les mêmes services techniques quel que soit leur environnement. Or, en fonction
des spécificités de l'environnement, certaines implantations d'un service sont plus efficaces que d'autres. Nous
proposons donc de réaliser également les \SNFs\ à l'aide du modèle à composants : un \SNF\ devient alors un
assemblage de composants élémentaires, que l'on peut remplacer pour faire évoluer le service. On associe à ces
nouveaux services techniques les notions de \pss\ et \pes\ qui facilitent leur description. Cela permet
l'adaptabilité statique ou dynamique des \SNFs. Afin d'utiliser les \pss\ et \pes, nous spécifions également un
ensemble de composants de gestion. Un prototype a été réalisé ayant pour but de démontrer la faisabilité d'un
tel système.
\section{\label{section2}Hétérogénéité et \QoS}
Le but de notre travail est d'apporter aux nouvelles applications distribuées, sur des environnements
hétérogènes, une \QoS\ dans n'importe quel contexte d'exécution, tout en maintenant leur interopérabilité, leur
réutilisabilité et leur simplicité d'implantation. En effet, l'émergence simultanée des outils informatiques
personnels portables et des réseaux sans fil a considérablement augmenté le nombre et les types des terminaux
connectés aux réseaux : serveurs puissants,  ordinateurs portables (PDA, téléphones mobiles) qui, bien que
connectés au réseau et offrant des capacités de stockage et de traitement, sont bien moins puissants, terminaux
de télévision numérique, consoles de jeux. De plus, des réseaux pair à pair remplaçant de plus en plus souvent
les architectures classiques client/serveur, les participants de l'application passent successivement du rôle de
client à celui de serveur, pour fournir des donnée ou des services applicatifs, ceci quelques soient leurs
caractéristiques propres en terme de connexion et de puissance de calcul. Pour développer des applications
portables sur ces terminaux, il est donc nécessaire de prendre en compte plusieurs contraintes liées à leur
hétérogénéité. Pour le code applicatif, des solutions d'adaptation existent, telles que CESURE \cite{CESURE} ou
ARCAD \cite{ARCAD} : elles reposent sur l'idée que le développeur d'application écrit plusieurs versions d'un
composant, chacune adaptée à son environnement puis décrit les adaptations. Ces solutions sont tout a fait
intéressantes mais augmentent fortement le travail du développeur. Nous proposons de compléter cette approche en
limitant le travail du développeur en adaptant automatiquement le code technique. L'intérêt étant que le code
technique, réutilisable pour de nombreuses applications, est développé par des spécialistes conscients des
particularités des environnements d'exécution. Comme pour les composants applicatifs, le code non-fonctionnel
doit bénéficier de tels mécanismes qui tiennent compte des spécificités de leur environnement d'exécution et du
contexte applicatif. Par exemple, pour les transactions \cite{her04}, il existe différents modèles théoriques de
gestion des transactions (plates, imbriquées, à flot de tâches). Chaque modèle transactionnel s'utilise dans des
contextes applicatifs bien précis, sur des terminaux ayant des capacités spécifiques : on ne fera pas de
transactions imbriquées sur une carte à microprocesseur. Ainsi, un bon nombre de \SNFs\ devront être adaptés aux
participants de l'application afin d'apporter la meilleure \QoS\ possible.
 Pour l'instant, avant l'exécution, le choix (d'utiliser ou non un service, le choix du modèle théorique ou
encore celui de l'implantation de ce modèle théorique) est difficilement possible, car les \SNFs\ dans les
plates-formes à composants industrielles sont figés lors du développement de la plate-forme. Afin de maintenir
l'interopérabilité, deux composants, même s'ils fonctionnent sur deux ORB différents, utilisent les mêmes
modèles de \SNFs. De plus, tout au long de l'exécution, il n'est plus possible de modifier le choix du modèle
théorique des \SNFs\ et de leur implantation. Il y est donc impossible de s'adapter aux changements éventuels de
l'environnement. Afin de permettre à un composant applicatif
 d'utiliser
n'importe quelle implantation du service transactionnel, \cite{rouvoy} propose une plate forme de déploiement:
en aval un adaptateur, spécifique à chaque implantation du service technique, fait l'interface entre le service
transactionnel et sa gestion par le conteneur.
\section{Intergiciels adaptables et modèle Fractal}
Nous avons vu qu'une plate-forme proposant une version unique des \SNFs\ n'est pas efficace dans les
environnements hétérogènes. On se propose donc de définir des mécanismes permettant de fournir plusieurs
implantations d'un même \SNF\ grâce à la notion d'intergiciel adaptable et plus particulièrement sur la
propriété de réflexivité du modèle à composants Fractal.
\subsection{\label{Intergiciels adaptables} Intergiciels adaptables} Après avoir été employé dans les langages de
programmation et les systèmes d'exploitation, l'utilisation de la réflexivité s'est largement répandue dans les
intergiciels. En effet, le besoin d'adaptation s'est fait ressentir dans de nombreux domaines de la
programmation distribuée (ex: en programmation orientée objet avec le projet Molène \cite{art65}). Néanmoins ces
solutions ne tirent pas partie des avantages provenant de la programmation par composants. Dans le domaine des
bus logiciels, il existe dynamicTAO \cite{dynamicTOA}, OpenORB \cite{OpenORB}, OpenCORBA \cite{OpenCORBA},
Flexinet \cite{Flexinet}. Ces intergiciels offrent un grand champ d'adaptations, cependant ils n'intègrent pas
l'idée d'un nombre indéterminé de \SNFs\ \cite{art31} et ne proposent pas véritablement de représentation
adaptée des \SNFs. Dans le domaine de la programmation par composants, on trouve le projet ARCAD \cite{ARCAD}
avec l'adaptation transparente des composants Fractal:  expérimentation visant à montrer qu'un service (pas
forcement technique), sous forme de composant Fractal, peut être dynamiquement adapté. Comme dans notre
solution, on utilise pour l'adaptation des règles ainsi que des informations sur son environnement. A la
différence de notre solution, le traitement de l'adaptation se fait dans le service lui-même, cela pourrait
poser des problèmes de cohérence entre l'adaptation de différents services. On peut aussi citer l'initiative
ArcticBeans dont la problématique est très proche de la nôtre. Cependant, ce projet se concentre sur deux
services (sécurité et transactions) et ne propose pas de solution générale. Enfin, le domaine de l'AOP a pour
but la réutilisation du code et se base sur le principe de séparation des préoccupations \cite{art39}. L'AOP
propose de décomposer les préoccupations transverses de l'application sous la forme d'aspects puis de les
recomposer dans une phase de tissage. L'AOP fournit des solutions statiques (ex : AspectJ \cite{AspectJ}) et
dynamiques (ex : les plates-formes de développement Prose \cite{Prose}, JAC \cite{JAC}, le serveur applicatif
JBoss \cite{JBoss}). L'AOP, notamment JAC, offre donc une solution pour réutiliser le code des services
technique et des techniques de tissage très complètes (ajout de code avant et après l'exécution d'une méthode
mais aussi lors de l'utilisation d'un objet). Cependant ne faisant pas de distinction catégorique entre service
technique et service réalisé par l'application, l'AOP ne propose donc pas pour l'instant de solution complète
pour gérer efficacement les services techniques.
\subsection{Fractal : réflexivité et assemblage de composants}
Le but du modèle à composants Fractal (\cite{fractal}, \cite{art40}) est d'offrir un cadre architectural global
pour le développement d'applications à base de composants. Contrairement à d'autres modèles tels que les EJB,
Avalon et CCM qui offrent un modèle concret du composant, Fractal regroupe une hiérarchie ou une famille de
modèles. On y distingue le modèle de traitement abstrait qui est la racine de la famille de modèle et définit
peu de concepts (composant, contrôleur, contenu, signal, nom et valeur). Un composant y est défini comme la
composition d'un contrôleur et d'un contenu, et le contrôleur d'un composant comme l'incarnation du comportement
de contrôle associé à ce composant. De plus, ce modèle définit un ensemble de propriétés de base du modèle dont
les propriétés d'encapsulation, d'abstraction et de récursion. Le modèle est complètement récursif et autorise
l'imbrication dynamique et le partage des composants à un niveau arbitraire. La hiérarchie des modèles de
Fractal étant extensible, il est possible de lui ajouter d'autres modèles (de traitements, de programmation ou
d'ingénierie). Deux des autres modèles que nous utiliserons pour décrire notre solution sont le modèle de
traitement concret et le modèle de programmation associé au framework Julia (i.e. l'implantation de référence de
Fractal) qui peut être utilisé pour la programmation de composants Fractal en Java. Dans ce modèle, le
contrôleur encapsulant le composant permet de contrôler la composition et les liaisons d'une composition de
composants. En effet, ces liaisons sont explicites et accessibles. Ainsi le système a une représentation de
lui-même. D'autre part, comparativement au conteneur des modèles à composants de métier (EJB, CCM), le
contrôleur proposé dans le modèle concret de Fractal propose peu de \SNFs\ (la gestion de cycle de vie). En
revanche, il laisse une totale liberté quant à l'ajout de \SNFs.  Néanmoins, il fournit des fonctionnalités
minimales permettant l'interception d'appel entrant ou sortant sur un composant, grâce aux notions
d'intercepteur et de sous-contrôleur. Un intercepteur peut être placé sur une interface client ou une interface
serveur, il "intercepte" les appels de méthodes et effectue des appels sur les sous-contrôleurs qui lui sont
associés. Les sous-contrôleurs fournissent différentes fonctionnalités d'introspection du composant (ex:
sous-contrôleur de contenu du composant ou de ses liaisons). Fractal laisse toute possibilité d'ajout de
nouveaux intercepteurs et sous-contrôleurs.
\section{\label{section5}Gestion des \SNFs\ adaptables}
Nous avons vu dans les sections précédentes que les \SNFs\ demandaient une plus grande flexibilité, pour être en
adéquation avec les environnements hétérogènes des applications distribuées. Or, les \SNFs\ des plate-formes
industrielles ne sont pas conçus sous forme de composants. Néanmoins, il est plus facile d'adapter ces services
aux besoins des applications en les définissant grâce au modèle à composants \cite{her04}. En effet, on peut
alors définir le \SNF\ comme étant un assemblage de plusieurs composants élémentaires. Pour proposer une version
adaptée d'un \SNF, le programmeur se contente de réaliser un assemblage des composants dont il a besoin. De
plus, le \SNF\ sous forme d'un composant alors
généré, "hérite" des caractéristiques positives d'un composant Fractal.\\
Dans \cite{her032}, nous avions aussi montré comment la définition d'un \SNF\ comme une composition de
composants permet d'accéder au modèle le plus adapté d'un ou plusieurs \SNFs, respectivement sous la forme d'une
\ps\ ou d'un \pe. Au niveau d'un seul service (ex: transaction, persistance), la \ps\ ("Personnalité d'un
Service") permet d'associer à un composant le modèle d'un \SNF\ le plus adapté; pour éviter que cette nouvelle
liberté dans le choix du modèle de \SNF\ n'entraîne une gestion trop lourde, nous représentons un ensemble de
\SNFs\ cohérent comme une entité appelée \pe\ ("Lot de Services"). Elle simplifie la gestion des \SNFs\ en la
résumant, du côté composant, au choix d'un \pe\ adapté à l'environnement d'exécution. De plus, elle garantit une
cohérence sémantique entre les \SNFs\ d'un même \pe.
\\Dans cette partie, nous définirons un ensemble de composants pour gérer la mise en place et la mise à jour de
telles compositions. Tout d'abord nous définirons l'utilité de chacun de ces composants de gestion. Puis nous
décrirons leur comportement au cours de l'exécution des applications. Un dernière partie sera consacrée plus
particulièrement à la définition de l'annuaire enregistrant les \SNFs.
\subsection{Définition des composants de gestion du système}
\begin{figure}[]
\begin{center}
\epsfig{width=7.2cm, file=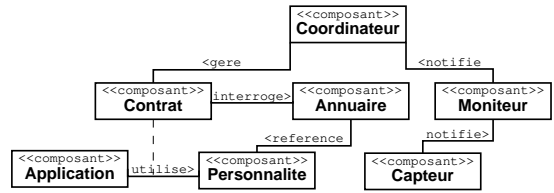} \caption{\label{figure system}Composants de
gestion du système}
\end{center}
\end{figure}
Afin de localiser, choisir, configurer et d'utiliser les personnalités \ps\ et \pe, nous définissons des
composants de gestion du système: des contrats, un coordinateur du système, un annuaire et des moniteurs (cf.
figure \ref{figure system}). Un {\it contrat} passé entre des personnalités et une application représente le
fait que cette application bénéficie des services de ces personnalités. De plus, il met à jour cette coopération
en trouvant la personnalité la plus adaptée par rapport à l'environnement d'exécution et aux besoins de
l'application. Le rôle du {\it coordinateur} est tout d'abord de gérer les contrats. Il les crée, les notifie
lorsqu'ils ont besoin d'une mise à jour et les détruit. Il reçoit les informations concernant l'environnement
grâce aux moniteurs auxquels il a souscrit en définissant ses préférences. Un {\it annuaire} de personnalités
fournit un service de nommage et courtage afin de trouver les personnalités des \SNFs. Il permet au contrat de
trouver une ou plusieurs personnalités en fonction de l'environnement d'exécution et des besoins de
l'application. Pour cela les personnalités doivent s'enregistrer à son service. Un {\it moniteur} reçoit des
informations de son capteur et fait une sélection en fonction des préférences que lui a exprimé le coordinateur.
Ainsi le coordinateur n'est pas surchargé par des informations inutiles. Ensuite il
envoie les informations pertinentes au coordinateur.\\
Durant le déploiement, l'administrateur installe les différents composants. A partir de ce moment, concernant
l'{\it adaptabilité statique}, il n'y a pas de différence avec une approche classique c'est-à-dire que le
composant applicatif bénéficie des services des personnalités qui ont été intégrées à sa composition.
\begin{figure}[]
\begin{center}
\epsfig{width=10.3cm, file=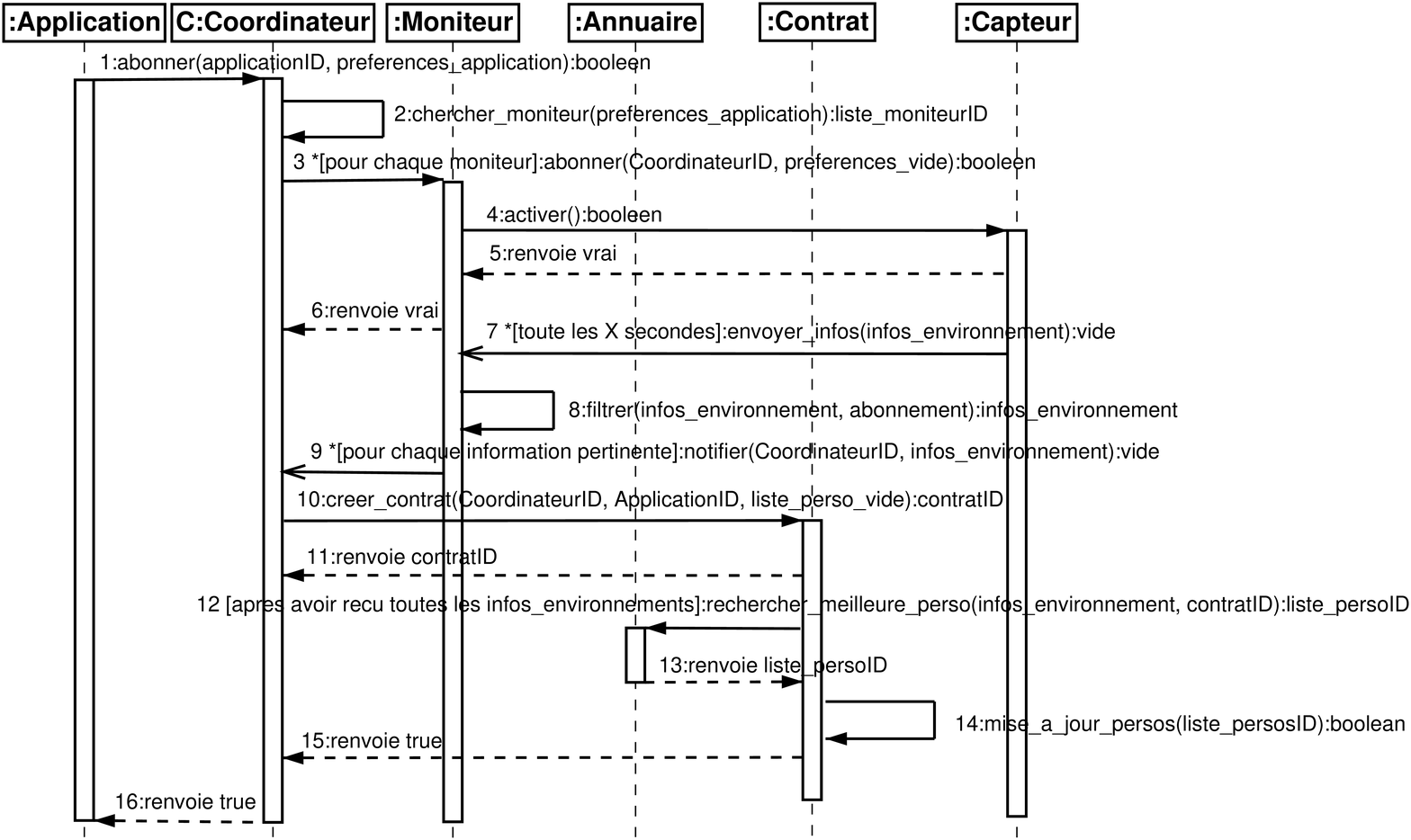} \caption{\label{contract_creation}Création d'un
contrat}
\end{center}
\end{figure}
Pour l'{\it auto-adaptabilité}, les composant applicatifs, l'annuaire et les moniteurs s'enregistrent auprès du
coordinateur. Les personnalités s'enregistrent dans l'annuaire. Les composants applicatifs fournissent au
coordinateur leurs préférences. Puis pour chaque composant applicatif, le coordinateur crée un contrat. Ce
dernier va chercher les personnalités adaptées à ajouter au composant applicatif (cf. figure
\ref{contract_creation}). Pour cela, il prévient le coordinateur de la liste des moniteurs qui sont intéressants
vis-à-vis des préférences de l'application. Par exemple, si l'application a besoin d'un haut niveau de sécurité,
il choisira un moniteur de réseau. Ensuite le coordinateur s'abonne auprès de chaque moniteur concerné qui est
alors activé. Seuls les moniteurs pertinents sont activés dans la mesure où il ne faut pas surcharger le
système. Le moniteur active les capteurs qu'il gère. Chaque capteur envoie alors des informations qui sont
filtrées par le moniteur. Le contrat peut alors créer les liaisons nécessaires entre le composant applicatif,
les personnalités et le contrôleur. Le composant applicatif peut maintenant s'exécuter avec les \SNFs\
appropriés à ses préférences et à son
environnement d'exécution.\\
Pendant l'exécution, l'environnement peut évoluer. Ainsi les contrats doivent être renégociés. Le système initie
cette adaptation grâce aux informations sur l'environnement qui lui sont données par les moniteurs. Le
coordinateur prévient les contrats qui sont concernés qu'ils doivent évaluer la nécessité de leur propre mise à
jour et chercher à nouveau les personnalités les plus adaptées grâce à l'annuaire.
\subsection{Focus sur l'annuaire de services techniques}
Nous développons un nouvel annuaire de services techniques, qui  permet à la plate-forme l'ajout et le retrait
dynamique de service ainsi que la gestion de multiples versions d'un même service. Il tient compte de
l'environnement d'exécution des services techniques qu'il référence afin de fournir au composant applicatif qui
le sollicite le service technique le plus adapté à ses besoins. En effet, grâce aux moniteurs (pour les
caractéristiques dynamiques) et aux informations données par l'administrateur (pour les caractéristiques
statiques), on connaît les capacités mémoire, réseau, l'architecture logiciel (type de bus logiciel,
environnement multi-processus), l'architecture matérielle (CPU, carte graphique), etc. Les services techniques
s'enregistrent dans cet annuaire en fournissant des renseignements sur l'environnement idéal pour lequel ils
sont adaptés (ex: l'implantation A du service transactionnel est adaptée à l'utilisation dans un réseau de type
Bluetooth c'est-à-dire, entre autre, avec de nombreuses déconnexions). Ces informations, ainsi que les
informations concernant les besoins des composants applicatifs constituent deux demi-contrats dans lesquels les
deux types de composants, applicatifs et techniques, expriment respectivement leurs besoins et les services
qu'ils offrent. Afin de fournir le service technique adéquat, l'annuaire vérifie que ces deux demi-contrats sont
bien compatibles pour former un contrat. Afin d'établir ce contrat, on fournit à l'annuaire un ensemble de
règles qui décrit les adaptations à effectuer. Ces règles peuvent être du type "l'implantation A du service de
tolérance aux fautes offre une plus grande qualité de service que l'implantation B". Ces règles seront notamment
établies par le développeur de services techniques, on laisse la possibilité à l'administrateur de la
plate-forme d'en ajouter de nouvelles. Afin de développer cet annuaire, nous nous reposons sur des technologies
déjà existantes telles que les annuaires de nommage et de courtage de CORBA et les annuaires LDAP. Les annuaires
déjà existants ne sont pas totalement satisfaisants pour résoudre notre problématique: dans CORBA des annuaires
de courtage permettent aux composants de retrouver d'autres composants. Les paramètres de leur requête étant le
nom et les paramètres d'exécution de ce service. Cependant, les services techniques n'étant pas eux même
considérés comme des composants, on ne peut pas les retrouver grâce à ce mécanisme. Notre démarche consiste à
dire que les services techniques sont des composants, on peut donc les enregistrer dans ce type d'annuaire. Même
ainsi ce type d'annuaire n'est pas spécialisé pour retrouver ces composants techniques, on ne peut donc pas
tirer partie de leurs particularités, des informations connues sur ces services techniques et surtout ils ne
gèrent pas d'information concernant l'environnement d'exécution.
\section{\label{section6}Prototype}
\begin{figure}[]
\begin{center}
\epsfig{width=6.1cm, file=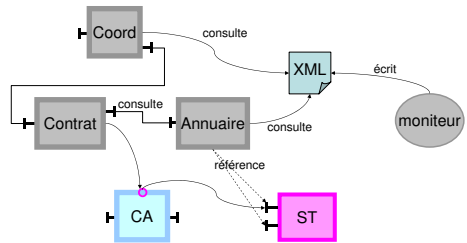} \caption{\label{composantsdegestion}Architecture du
prototype.}
\end{center}
\end{figure}

Cette partie est consacrée à notre prototype développé en Java à l'aide du modèle concret de Fractal nommé Julia
v 1.0. Notre système fonctionne de la façon suivante: on instancie et lie les composants de gestion comme décrit
dans la figure \ref{composantsdegestion}. Le coordinateur, l'annuaire et le contrat sont des composants (comme
bien entendu le composant applicatif et le service technique). Le moniteur est écrit en langage C et échange les
informations concernant l'environnement à travers un fichier XML. Nous en détaillons les raisons dans la partie
"le moniteur". Le contrat utilise une référence sur l'objet sous-contrôleur pour l'exécuter (référence
représenté par une flèche). Enfin l'annuaire référence les interfaces des \SNFs\ (référence représenté par une
flèche en pointillé). C'est cette référence qui sera donnée au contrat comme résultat d'une requête auprès de
l'annuaire.
les composants de gestion sont:
\begin{itemize}
\item Le {\it coordinateur} joue le rôle d'usine de contrat. En effet, lorsqu'un nouveau composant applicatif
doit être ajouté à l'application, l'administrateur contacte le coordinateur qui fournit un contrat qui est dès
lors associé au composant applicatif. Il joue aussi le rôle de déclencheur de mise à jour du système lorsqu'il
détecte une modification de l'environnement d'exécution. \item Le {\it contrat} gère la liaison entre le
composant applicatifs et les \SNFs\ ainsi que l'interrogation de l'annuaire. Lorsqu'il repère qu'une \SNF\ n'est
plus adapté à l'environnement, il interroge l'annuaire pour en trouver un nouveau. Nous commençons par
développer une solution qui implante la fonctionnalité principale du composant contrat: la création d'une
composition telle que décrite dans \cite{her04}. Pour le coordinateur prend pour paramètres le composant
applicatif (qui peut être une composition de composants) et un fichier décrivant l'ensemble des \SNFs\ dont il a
besoin (leur nom, ex: persistance, transaction, etc.) et pour chaque \SNF, la \ps\ que l'on veut utiliser (une
\ps\ étant décrite ici par le nom de l'interface correspondante). \item Afin d'écrire le {\it moniteur}, nous
avons évalué différentes solutions. L'ensemble du système étant écrit avec Julia, nous avons cherché s'il était
possible d'écrire un moniteur en Java afin de l'encapsuler dans un composant Fractal, portable sur de nombreuses
machines. Cependant, le langage Java, dans la mesure où il n'a pas été conçu dans cette optique, ne permet pas
d'accéder à des données de bas niveau. Nous avons aussi évaluer des solutions tels que l'utilisation du logiciel
Aida32 \cite{aida32}, WMI (Windows Management Instrumentation) pour Windows \cite{WMI} ou encore rechercher des
informations dans les répertoires de "/proc" sur Linux mais ces solutions sont dépendantes du système
d'exploitation et ne nous donnent pas tous les renseignements nécessaires. Nos recherches se sont donc tournées
vers le langage C. Nous avons donc développé un moniteur en C qui est recompilé suivant l'environnement
d'exécution. Pour faire l'interface entre le coordinateur et le moniteur, nous avions alors le choix entre une
interface JNI (Java Native Interface) et l'échange de données à travers un fichier. La première solution étant
plus complexe à mettre en oeuvre et n'ayant pas plus d'intérêt que la seconde, nous avons opté pour la seconde.
Ce moniteur récupère donc un ensemble de données (réseau, CPU, logiciel, etc) puis les inscrit dans un fichier
XML tel que celui de la figure \ref{exempleXMLenvironnement}.
\begin{figure}[]
\begin{center}
\epsfig{width=6.5cm, file=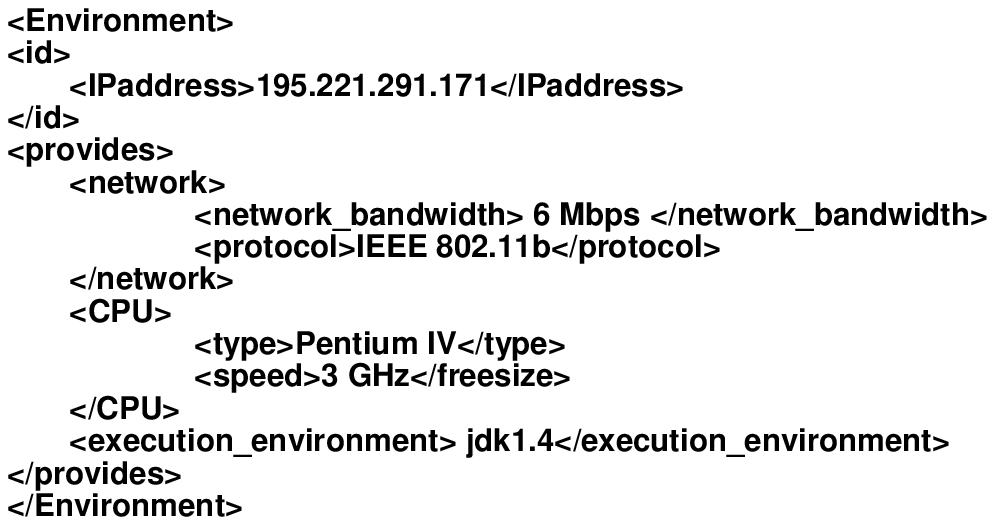} \caption{\label{exempleXMLenvironnement}Exemple de
description de l'environnement fournie par le moniteur.}
\end{center}
\end{figure}
\item L'{\it annuaire} fournit une interface regroupant des opérations de d'import et d'export. Les opérations
d'import sont effectuées par l'administrateur lors de l'ajout de services techniques à la plate-forme. Les
opérations d'export sont faites par l'annuaire lorsqu'il consulte l'annuaire. Pour ces deux types d'opérations,
on doit répondre à deux questions: comment stocker les données dans notre annuaire et répondre à une requête.
Nous avons utilisé un annuaire de composants développé au sein de notre équipe pour Fractal et en Fractal puis
nous l'avons spécialisé. Cet annuaire de base est une adaptation du concept d'annuaire CORBA dans le modèle
Fractal. Cependant, certaines modifications du concepts ont été effectuées pour être en adéquation avec le
modèle Fractal telles que le référencement des patrons et de types. En effet, Dans le modèle CORBA seules les
instances de composants sont référencées dans l'annuaire, les patrons de composants étant stockés dans
l'"interface repository". Cela permet à notre annuaire d'instancier un patron s'il juge que le composant ainsi
instancié répondrait mieux aux attentes du système. Dans une seconde phase, nous avons spécialisé cet annuaire
pour les services techniques. La structure de stockage adoptée est un arbre dans lequel une instance de service
technique est le fils d'un patron et un patron le fils d'un type. Pour chaque type et chaque patron, on stocke
un identifiant ainsi que l'ADL correspondant. Pour chaque instance de service technique, on stocke un
identifiant, une référence. Aux patrons et aux instances, on associe un fichier XML décrivant le service fourni
(ex: un service transactionnel, fournissant les transactions plates, version 1.3), ensuite on y détaille les
besoins de \SNF\ en terme de réseau, environnement d'exécution, etc.
\end{itemize}
\section{\label{section8}Conclusion}
Cet article propose d'adapter des \SNFs, pour permettre le développement des architectures à composants dans de
nombreux domaines tels que le M-commerce où les applications s'exécutent dans des environnements divers (pour
l'adaptabilité statique) et très changeants (pour l'auto-adaptabilité). Nous avons rappelé la proposition d'une
adaptabilité à deux niveaux: au niveau d'un seul \SNF, représentée par la notion de \ps. Elle permet d'offrir
aux composants applicatifs le modèle du service le plus adapté à son environnement tout en maintenant
l'interopérabilité au sein du système. L'adaptabilité au second niveau se traduit par le concept de \pe, qui
permet de fournir un ensemble de \SNFs\ cohérent. Cela facilite également la gestion des problèmes
d'entrelacement des \SNFs. Afin de gérer les \pss\ et les \pes, nous proposons de fournir plusieurs entités: un
annuaire (pour retrouver les \SNFs), un contrat (pour représenter l'utilisation par une application d'une
personnalité), un coordinateur (pour créer et mettre à jour les contrats) et des moniteurs qui donnent au
coordinateur des informations concernant l'environnement d'exécution. Nous avons démontré la faisabilité des
versions statiques et dynamiques de l'adaptabilité grâce à un prototype basé sur l'implantation Julia du modèle
 Fractal. Concernant l'adaptabilité statique, nous montrons même que l'utilisation du modèle à
composants pour implanter les services techniques et l'utilisation des sous-contrôleurs n'entraînent
pratiquement aucun surcoût en terme de vitesse d'exécution (moins de 0,5\%). Nous avons rencontré quelques
problèmes techniques pour la réalisation de ce prototype (définition au préalable des sous-contrôleurs, pas
d'interface interne client pour un sous-contrôleur), mais le résultat obtenu reste proche de la solution
théorique envisagée et surtout il permettra une évolution facile vers la solution théorique si les prochaines
versions de Julia le permettent. Les bénéfices d'une telle solution sont l'amélioration de la \QoS\ et la
réutilisabilité du code tout en maintenant les caractéristiques des composants telles que l'interopérabilité et
la facilité d'implantation. Des problèmes subsistent, tels que la séparation des \SNFs\ (même si cela est en
partie résolue par le \pe) ou la description de la \QoS\ fournie par un composant. Cet article donne une
définition de l'utilisation des composants de gestion des \SNFs. Des travaux sont en cours pour rendre plus
performant chacun de ces composants. Pour cela, les règles de fonctionnement du coordinateur restent à
formaliser. Les modifications du contrôleur dues au contrat sont également à préciser. Cela fait l'objet de
travaux en cours. Concernant les moniteurs, on étudie les ressources supplémentaires à surveiller et comment le
faire. Dans nos travaux futurs, on s'intéressera certainement à l'AOP. En effet, dans la mesure où certains des
buts fixés par l'AOP et les nouveaux challenges auxquels la programmation par composants a à faire face sont les
mêmes, on peut imaginer une certaine convergence entre les deux domaines, l'AOP pouvant tirer partie des
recherches sur les composants ayant attrait à la distribution, l'interopérabilité et l'hétérogénéité et la
programmation par composants pouvant s'inspirer des techniques évoluées de composition des aspects proposées par
l'AOP.
\bibliography{DECOR_04}

\begin{thebibliography}{}

\bibitem[aid]{aida32}
\guilo{}http://www.aida32.hu/\guilf{}.

\bibitem[AND~99]{art65}
\textsc{André F.}\andname{}\textsc{Segarra M.-T.}, \guilo{}A Generic Approach
  to Build Mobile Applications.\guilf{},
\newblock \technicalreportname{} \numbername 3723, Juin 1999, INRIA.

\bibitem[BLA~01]{OpenORB}
\textsc{Blair G.}, \textsc{Coulson G.}, \textsc{Andersen A.}, \textsc{Blair
  L.}, \textsc{Clarke M.}, \textsc{Costa F.}, \textsc{Duran-Limon H.},
  \textsc{Fitzpatrick T.}, \textsc{Johnston L.}, \textsc{Moreira R.},
  \textsc{Parla-vantzas N.}\andname{}\textsc{Saikoski K.}, \guilo{}The design
  and implementation of Open ORB 2.\guilf{},
\newblock \Inname{} \textit{IEEE Distrib. Syst. Online 2}, IEEE, Septembre
  2001.

\bibitem[BRU~02]{art40}
\textsc{Bruneton E.}, \textsc{Coupaye T.}\andname{}\textsc{Stefani J.-B.},
  \guilo{}Recursive and Dynamic Software Composition with Sharing\guilf{},
\newblock \Inname{} \textit{Proceedings of the 7th ECOOP International Workshop
  on Component-Oriented Programming (WCOP'02)}, Malaga, Espagne, 2002.

\bibitem[DUC~02]{art31}
\textsc{Duclos F.}, \guilo{}Environnement de Gestion de Services
  Non-Fonctionnels dans les Applications à Composants\guilf{},
\newblock Thèse de doctorat, Université J. Fourier de Grenoble, 2002.

\bibitem[fra]{fractal}


\bibitem[HAY~97]{Flexinet}
\textsc{Hayton R.}, \guilo{}FlexiNet Open ORB Framework.\guilf{},
\newblock \technicalreportname{} \numbername 2047.01.00., 1997, APM ltd,
  Poseidon House, Castle Park, Cambridge, UK.

\bibitem[HéR~03]{her032}
\textsc{Hérault C.}\andname{}\textsc{Lecomte S.}, \guilo{}Adaptabilité des
  Services Techniques dans un Modèle à Composants\guilf{},
\newblock \Inname{} \textit{3ème Conférence Française sur les Systèmes
  d'Exploitation (CFSE)}, La Colle sur Loup, France, Octobre 2003.

\bibitem[HéR~04]{her04}
\textsc{Hérault C.}, \textsc{Nemchenko S.}\andname{}\textsc{Lecomte S.},
  \guilo{}A Component-Based Transactional Service, Including Advanced
  Transactional Models\guilf{},
\newblock \Inname{} \textit{Fourth IEEE International Symposium and School on
  Advance Distributed Systems (ISSADS 04)}, Guadalajara, Jalisco, Mexique,
  Janvier 2004.

\bibitem[JBo]{JBoss}
\guilo{}http://www.jboss.org\guilf{}.

\bibitem[KIC~97]{art39}
\textsc{Kiczales G.}, \textsc{Lamping J.}, \textsc{Mendhekar A.}, \textsc{Maeda
  C.}, \textsc{Lopes C.~V.}, \textsc{Loingtier J.-M.}\andname{}\textsc{Irwin
  J.}, \guilo{}Aspect-Oriented Programming\guilf{},
\newblock \Inname{} \textit{Proceedings of ECOOP}, Springer-Verlag, 1997.

\bibitem[KIC~01]{AspectJ}
\textsc{Kiczales G.}, \textsc{Hilsdale E.}, \textsc{Hugunin J.},
  \textsc{Kersten M.}, \textsc{Palm J.}\andname{}\textsc{Griswold W.},
  \guilo{}Getting started with ASPECTJ\guilf{},
\newblock \textit{Communications of the ACM}, \volumename\ 44, \numbername\ 10,
  2001,  \pagesname{} 59--65, ACM Press.

\bibitem[KON~00]{dynamicTOA}
\textsc{Kon F.}, \textsc{Roman M.}, \textsc{Liu P.}, \textsc{Mao J.},
  \textsc{Yamane T.}, \textsc{Magalhaes L.}\andname{}\textsc{Campbell R.},
  \guilo{}Monitoring, security, and dynamic configuration with the dynamicTAO
  reflective ORB.\guilf{},
\newblock \Inname{} \textit{Proceedings of the IFIP/ACM International
  Conference on Distributed Systems Platforms and Open DistributedProcessing
  (Middleware2000)}, Palisades, NY, Avril 2000, Springer-Verlag,  \pagesname{}
  121-143.

\bibitem[LAV~01]{WMI}
\textsc{Lavy M.}\andname{}\textsc{Meggitt A.}, \guilo{}Windows Management
  Instrumentation (WMI)\guilf{},
\newblock \technicalreportname{} \numbername ISBN: 1578702607, Octobre 2001,
  New Riders Publishing.

\bibitem[LED~99]{OpenCORBA}
\textsc{Ledoux T.}, \guilo{}OpenCORBA: A reflective open broker.\guilf{},
\newblock \Inname{} \textit{Proceedings of Reflection'99}, Saint-Malo, France,
  Juillet 1999, Springer-Verlag,  \pagesname{} 197-214.

\bibitem[PAW~03]{JAC}
\textsc{Pawlak R.}, \textsc{Seinturier L.}\andname{}\textsc{Duchien L.},
  \guilo{}Jac Milestone 2003\guilf{},
\newblock \technicalreportname{} \numbername 2003-4, 2003, LIFL, Université des
  Sciences et Technologies de Lille.

\bibitem[PEL~00]{CESURE}
\textsc{Pellegrini M.-C.}, \textsc{Potonniée O.}, \textsc{Marvie R.},
  \textsc{Jean S.}\andname{}\textsc{Riveill M.}, \guilo{}CESURE : une
  plate-forme d'applications adaptables et sécurisées pour usagers
  mobiles\guilf{},
\newblock \Inname{} \textit{Special Issue of the french journal Calculateurs
  Parallèles, entitled "Évolution des plates-formes orientées objets
  répartis"}, Hermès, 2000,  \pagesname{} 113-120.

\bibitem[Pro]{Prose}
\guilo{}http://prose.ethz.ch\guilf{}.

\bibitem[PRO~03]{ARCAD}
\textsc{Projet RNTL~ARCAD c. T.~L.}, \guilo{}D1.3 Document
  d'architecture\guilf{},
\newblock \technicalreportname{}, 2003.

\bibitem[RIV~02]{art30}
\textsc{Riveill M.},
\newblock \guilo{}Programmation par composition et déploiement d'applications
  réparties\guilf{},
\newblock Coopération dans les systèmes à objets (RSTI-Série L'Objet Vol.8 N°
  3), 2002.

\bibitem[ROU~03]{rouvoy}
\textsc{Rouvoy R.}\andname{}\textsc{Merle P.}, \guilo{}Abstraction of
  Transaction Demarcation in Component-Oriented Platforms\guilf{},
\newblock \Inname{} \textit{ACM/IFIP/USENIX International Middleware Conference
  (Middleware'03)}, Rio de Janeiro, Brésil, Juin 2003.

\end{thebibliography}
\end{document}